\documentclass[12pt]{iopart}
\usepackage{amssymb}
\usepackage{graphics}
\usepackage{graphicx}
\usepackage[usenames]{xcolor}
\usepackage{bm}

\def\secref#1{section~\ref{#1}}

\def\fig#1{Figure~\ref{#1}}

\def\eqref#1{Eq.~(\ref{#1})}

\def\eq#1{equation~(\ref{#1})}

\def\bea{\begin{eqnarray}}
\def\eea{\end{eqnarray}}

\def\be{\begin{equation}}
\def\ee{\end{equation}}

\def\({\left(}
\def\){\right)}
\def\[{\left[}
\def\]{\right]}

\def\D'{\Delta'}

\def\gmax{\gamma_{\rm max}}
\def\kmax{k_{\rm max}}

\def\etal{\textit{et al.}}

\begin{document}
\title[Magnetic reconnection: from the Sweet-Parker model to stochastic plasmoid chains]{Magnetic reconnection: from the Sweet-Parker model to stochastic plasmoid chains}
\author{N F Loureiro$^1$, D A Uzdensky$^2$}
\address{$^1$ Instituto de Plasmas e Fus\~ao Nuclear, Instituto Superior T\'ecnico, 
Universidade de Lisboa, 1049-001 Lisbon, Portugal}
\address{$^2$ Center for Integrated Plasma Studies, Physics Department, 
UCB-390, University of Colorado, Boulder, Colorado 80309, USA}
\ead{nloureiro@ipfn.ist.utl.pt}
\begin{abstract}
Magnetic reconnection is the topological reconfiguration of the magnetic field in a plasma, accompanied by the violent release of energy and particle acceleration. Reconnection is as ubiquitous as plasmas themselves, with solar flares perhaps the most popular example. Other fascinating processes where reconnection plays a key role include the magnetic dynamo, geomagnetic storms and the sawtooth crash in tokamaks.

Over the last few years, the theoretical understanding of magnetic reconnection in large-scale fluid systems has undergone a major paradigm shift. The steady-state model of reconnection described by the famous Sweet-Parker (SP) theory, which dominated the field for $\sim 50$ years, has been replaced with an essentially time-dependent, bursty picture of the reconnection layer, dominated by the continuous formation and ejection of multiple secondary islands (plasmoids). 
Whereas in the SP model reconnection was predicted to be slow, a major implication of this new paradigm is that reconnection in fluid systems is fast (i.e., independent of the Lundquist number), provided that the system is large enough.
This conceptual shift hinges on the realization that SP-like current layers are violently unstable to the plasmoid (tearing) instability --- implying, therefore, that such current sheets are super-critically unstable and thus can never form in the first place. This suggests that the formation of a current sheet and the subsequent reconnection process cannot be decoupled, as is commonly assumed.

This paper provides an introductory-level overview of the recent developments in reconnection theory and simulations that led to this essentially new framework. We briefly discuss the role played by the plasmoid instability in selected applications, and describe 
some of the outstanding challenges that remain at the frontier of this subject.
Amongst these are the analytical and numerical extension of the plasmoid instability to (i) 3D and (ii) non-MHD regimes. New results are reported in both cases.
\end{abstract}

\submitto{\PPCF}


\maketitle

\section{Introduction}
\label{sec:intro}

\subsection{Statement of the problem}
Magnetic reconnection is the change in magnetic-field topology that enables phenomena such as solar flares and coronal mass ejections, substorms in the Earth's magnetic field and the sawtooth and tearing instabilities in magnetic-confinement fusion~\cite{biskamp_magnetic_2000,zweibel_magnetic_2009,yamada_magnetic_2010}. 
Any such topological reconfiguration requires the breaking of the frozen-flux constraint of ideal MHD. 
Understanding the detailed physics of how this happens, and the ensuing energetics and dynamics of the plasma as reconnection occurs, is the goal of the investigations in this topic.

An important figure of merit in the characterisation of reconnection environments is the Lundquist number, $S=L V_A/\chi_m$, where $L$ is the scale length characteristic of the reconnecting field $\bm B_0$ (typically identified with the system size), $V_A=B_0/\sqrt{4\pi\rho}$ is the corresponding Alfv\'en speed ($\rho$ is the plasma's mass density) and $\chi_m= c^2/(4\pi\sigma)$ is the magnetic diffusivity, with $\sigma$ the electrical conductivity of the plasma. 
In typical solar coronal conditions, $S\sim 10^{12}-10^{14}$; in the Earth's magnetotail, $S\sim 10^{15}-10^{16}$; in a modern tokamak such as JET, $S\sim 10^6 - 10^8$. 
The fact that these are rather large numbers means that indeed frozen flux ought to be a very good approximation; 
the common understanding is that the occurrence of magnetic reconnection thus requires the formation of highly localised regions of very intense currents --- current sheets or, in some models, $X$-points --- where nonideal effects can (and must) become important. The simplest of such effects is the plasma resistivity; others are electron inertia and finite electron Larmor radius terms.

A wide variety of observations of reconnection events 
are characterised by three fundamental aspects: 
(i) {\it Fast reconnection} --- the rates of energy release are in the  range $\sim 0.01 - 0.1 V_A/L$  for most systems, seemingly independently of the magnitude of the frozen-flux breaking terms. 
(ii) {\it Two timescales} --- the reconnection stage proper (explosive) is preceded by a much slower stage of energy accumulation; the transition between the two stages is generally referred to as the reconnection {\it trigger}.
(iii) {\it Efficient energy conversion} --- magnetic reconnection is, essentially, a mechanism for efficiently converting magnetic energy of the reconnecting fields into bulk plasma heating, supra-thermal particle acceleration, and kinetic energy of the reconnection outflows.

A complete understanding of magnetic reconnection perforce must describe these three different aspects. However, despite roughly sixty years of very active research, one is forced to admit that such an understanding continues to elude us in even the simplest plasma description that allows for reconnection (resistive MHD). This paper is an attempt at overviewing a recent development in the field --- the plasmoid instability and ensuing plasmoid-dominated reconnection --- that, we believe, has brought us closer to a solution. 

Motivated by the ubiquity of plasmoids in magnetic reconnection, and given the wide variety of phenomena in which reconnection itself plays a fundamental role, we aim in this paper to  provide an accessible discussion of the plasmoid instability and dynamics, and some associated developments, to non-expert readers. 
It is, however, {\it not} a review paper of plasmoid-mediated reconnection; the selection of topics that we cover is guided by our personal interests and is manifestly incomplete.

In addition, this paper also includes some new results: \secref{sec:3D} reports preliminary studies of the plasmoid instability in 3D geometries, and, in \secref{sec:kin_plasmoids}, we present an analytical extension of the plasmoid instability to two new semi-kinetic regimes which we think should be of direct applicability to laboratory reconnection experiments as well as to the solar corona.

\subsection{The Sweet-Parker model}
The first serious attempt at describing magnetic reconnection was due to Peter Sweet~\cite{sweet_neutral_1958} and Eugene Parker~\cite{parker_sweets_1957}, giving rise to what came to be known as the Sweet-Parker (SP) model. 
They conceptualized a current sheet as a {\it steady-state} channel of length $L$ and thickness $\delta_{SP}$, through which plasma flows in incompressible fashion, with velocity $u_{in}$ upstream and $u_{out}$ downstream. 
Using simple dimensional arguments, Sweet and Parker were then able to show from the resistive MHD equations that (i) $u_{out}\sim V_A$, (ii) $u_{in}/u_{out}\sim S^{-1/2}$, (iii) $\delta_{SP}/L\sim S^{-1/2}$. These relationships imply that the electric field set up by the reconnection process, i.e., the rate of change of the magnetic flux, is $c E = V_A B_0 S^{-1/2}$, or, equivalently, that the reconnection rate, $\tau_{rec}^{-1}=u_{in}/L$, is given by $\tau_A/\tau_{rec}\sim S^{-1/2}$, where $\tau_A=L/V_A$ is the Alfv\'en time.

Given the rather large values of the Lundquist number that one tends to encounter in nature, the $S^{-1/2}$ scaling predicted by the SP model is clearly insufficient to explain the observed reconnection rates. For example, in typical solar-corona conditions, $\tau_A\approx 0.5s$, 
leading to the prediction that a typical solar flare should last $\sim2$ months, in stark contrast with the observed duration of $15$ minutes to $\sim 1$ hour. Similarly unsatisfactory predictions are obtained for reconnection events in almost all plasmas that one cares to examine.

\subsection{Petschek's solution}

The SP model's inability to yield faster reconnection rates stems directly from the very large aspect ratio of the current sheet that this model predicts.
An important attempt to circunvent this difficulty is due to Harry Petschek in 1964~\cite{petschek_magnetic_1964} (later revisited and amended by Russell Kulsrud~\cite{kulsrud_magnetic_2001}). 
Petschek's solution relies on shortening the length of the current sheet at the expense of four standing slow-mode shocks emanating from a central diffusion region. 
His model yields only a logarithmic dependence of the reconnection rate on $S$, i.e., Petschek reconnection is fast (in some cases, it is in fact too fast). 

Although this was encouraging, the absence of a justification for the origin of the shocks, and other non-rigorous assumptions invoked in the derivation, place the Petschek model on somewhat less firm grounds than the SP model.
And indeed, direct numerical MHD simulations of reconnection have failed to exhibit Petschek's solution~\cite{park_reconnection_1984,biskamp_magnetic_1986,uzdensky_two-dimensional_2000, loureiro_x-point_2005, malyshkin_magnetic_2005} even if it is used as the initial condition~\cite{uzdensky_two-dimensional_2000} -- instead, those simulations, and many others at moderately large values of the Lundquist number, $S\lesssim 10^4$, reproduce rather well all features of the SP model, as long as uniform, or smoothly varying (e.g, Spitzer), resistivities are employed --- a conclusion that has been corroborated by dedicated experiments~\cite{ji_experimental_1998}.
A noteworthy exception to this statement arises when strongly localized resistivity profiles are used (motivated by attempts to incorporate kinetic effects into the MHD description, e.g., kinetic-scale instabilities that may give rise to micro-turbulence localised to the current sheet and thus anomalously enhance the resistivity there~\cite{kulsrud_plasma_2005,buchner_vlasov_2005,buchner_anomalous_2006}). 
Then, numerical simulations do exhibit Petschek-like configurations~\cite{sato_externally_1979, ugai_computer_1995, scholer_undriven_1989,erkaev_reconnection_2000,erkaev_rate_2001,biskamp_localization_2001,malyshkin_magnetic_2005}. 
One question that then arises, of course, is how well-justified such anomalous resistivity models are; addressing this concern, however, requires (potentially 3D) fully kinetic simulations capable of reproducing MHD length-scales, a feat which is beyond current computational capabilities.

A more fundamental question that follows is whether fast reconnection is excluded from a pure MHD description. Addressing this question is the central aim of this paper.

\subsection{Sweet-Parker revisited}
The absence of solid numerical support for the Petschek model as a valid MHD solution prompts us to reassess the assumptions on which the SP model is based, and ask whether these are expected to hold in the reconnection environments typically found in nature. Perhaps first and foremost is the validity of MHD itself: for example, in the magnetotail collisions are so rare that kinetic scale effects are bound to become important, rendering MHD insufficient. Similar observations can be made for reconnection in modern magnetic fusion devices and a wide-variety of astrophysical and space environments~\cite{ji_phase_2011}. But there are certainly physical contexts where one expects MHD to hold and, simultaneously, either observes or infers that fast reconnection must be happening~\cite{loureiro_turbulent_2009,ji_phase_2011}. Can the SP model be wrong, or somehow inapplicable?

Two major assumptions invoked in the derivation of the SP model are that (i) {\it the background plasma is laminar}, and (ii) {\it a steady-state solution is realisable in practice, or, in other words, that the reconnection geometry central to the SP analysis (the current sheet) is stable}. Do these assumptions hold?

\subsubsection{Background turbulence.}
Background turbulence is indubitably present in many plasmas where reconnection takes place~\cite{karimabadi_magnetic_2013}, and one may legitemately wonder if its presence significantly affects reconnection. In particular, how does background turbulence change the predictions of the SP model?

This question was first raised in the pioneering numerical investigations of Matthaeus and Lamkin in 1986~\cite{matthaeus_turbulent_1986}. This study revealed many features strongly suggestive of the inadequacy of a steady-state analysis of the reconnection layer but, given the very limited numerical resolution then available, it was unclear whether turbulence could significantly speed up reconnection.

Several years later, the influence of background turbulence on reconnection was analysed theoretically in the landmark paper of Lazarian and Vishniac~\cite{LV_99} (hereafter LV99). The main conclusion of this work was that, in the presence of background turbulence, MHD reconnection should become fast, independent of $S$, but that 3D effects were crucial to achieve this (which would have precluded Matthaeus and Lamkin~\cite{matthaeus_turbulent_1986} from observing an enhancement of the reconnection rate due to turbulence because their simulations were 2D). 

The subsequent advent of massively parallel computing enabled the first attempt at the direct numerical testing of the LV99 model~\cite{kowal_numerical_2009} in 3D simulations. 
The limits set by the available computing power dictated a maximum value of the Lundquist number $S=2000$, not sufficiently large to allow the desirable asymptotic scale separation between the system size (set by the large-scale reconnecting magnetic field), the energy injection scale (i.e., the turbulence forcing scale), the width of the initial current sheet (a Sweet-Parker sheet) and the resistive and viscous turbulence dissipation scales. Modulo these constraints, Ref.~\cite{kowal_numerical_2009} reported a confirmation of several predictions of LV99, including, importantly, the increase in the reconnection rate to $S$-independent values.

Simultaneously, Ref.~\cite{loureiro_turbulent_2009} undertook similar simulations but only in 2D, at values of the Lundquist number ranging up to $S\approx 1.5\times 10^4$. 
Interestingly, they also reported fast reconnection; however, as mentioned above, such enhancement of the reconnection rate over the nominal SP value could not be attributed to the LV99 mechanism given its 2D nature.  Ref.~\cite{loureiro_turbulent_2009} conjectured instead that the reconnection speed-up that they observed may in fact be due to the breaking of the second assumption key to the SP model: rather than being steady-state, the current sheet was actually strongly unstable to the formation of multiple plasmoids~\cite{loureiro_instability_2007, loureiro_plasmoid_2013}, and the role of turbulence in this case was to facilitate the onset of this instability. 

\subsubsection{Instability of the current sheet.}
Numerical evidence for the instability of Sweet-Parker-like current sheets to secondary-island (plasmoid) formation was reported in the literature at least as early as 1984~\cite{steinolfson_nonlinear_1984, park_reconnection_1984}, and reiterated in subsequent numerical investigations~\cite{biskamp_magnetic_1986,lee_multiple_1986,jin_twodimensional_1991,ugai_computer_1995,loureiro_x-point_2005}.  Heuristic arguments put forth by Biskamp~\cite{biskamp_magnetic_1986}, based on earlier work by Bulanov \etal~\cite{bulanov_tearing-mode_1979}, suggested that Sweet-Parker current sheets of aspect ratio exceeding $\sim100$ (corresponding to $S\approx 10^4$) would always be unstable to plasmoid formation. This argument thus hinted at the existence of a critical value of the Lundquist number, $S_c$, below which the SP model applies, and above which an instability appears. 
But the question of what happens in the asymptotic limit $S\gg S_c$ remained open: due to the limited computational power available, those simulations could only marginally exceed $S_c$, and no link could be established between the instability of the sheet and the reconnection rate. 
The nature of the instability itself was also unclear --- what was the fastest growing unstable mode and the corresponding growth rate at asymptotically large values of the Lundquist number?

A separate, powerful, motivation to understand the structural stability of current sheets was also provided by kinetic (PIC) simulations of reconnection, where plasmoid formation is observed and conjectured to play an important, perhaps critical, role in the process, from influencing its rate and conferring reconnection a bursty, rather than steady-state, character~\cite{daughton_fully_2006,drake_formation_2006}, to enhancing the particle acceleration efficiency~\cite{drake_electron_2006}. Naturally, the stability of such fully kinetic current sheets (which is still not analytically understood, but see discussion in Ref.~\cite{liu_dispersive_2014})  is determined by more complex physics than is the case for the MHD ones, but one may hope to at least gain qualitative insight from a better understanding of the fluid case.

\section{The plasmoid instability}
\label{sec:plasmoid_inst}

The arguments put forth in the previous section motivated a concerted analytical and numerical effort to investigate rigorously the stability of SP-like current sheets at high values of the Lundquist number. 
In summary, such current sheets have been found to be violently unstable to the formation of multiple plasmoids, in what has come to be known as the plasmoid instability (but can, in fact, be correctly thought of as the large $\Delta'$ tearing instability~\cite{furth_finite-resistivity_1963,coppi_resistive_1976} of an equilibrium --- the SP current sheet --- whose characteristic scale is itself a function of resistivity: $\delta_{SP}\sim L S^{-1/2}$).
It is now generally accepted that this instability radically changes MHD reconnection from the Sweet-Parker steady-state picture: plasmoid-mediated reconnection is intrinsically non-steady-state, bursty and fast. 
The main aspects of the linear and nonlinear stages of the plasmoid instability are reviewed next.

\subsection{Linear stage}
\label{sec:linear_stage}

To compute analytically the linear instability of resistive-MHD, incompressible, SP current sheets, Loureiro \etal~\cite{loureiro_instability_2007} resorted to standard tearing-mode techniques~\cite{furth_finite-resistivity_1963,coppi_resistive_1976}.
They found a very violent instability, whose fastest growing mode had a wavenumber $\kmax L\sim S^{3/8}$ and corresponding growth rate of $\gmax\tau_A\sim S^{1/4}$; the width of the corresponding resistive boundary layer inside the (equilibrium) SP sheet on which the plasmoids are born is $\delta_{in}/\delta_{SP}\sim S^{-1/8}$. These scalings for the wavenumber and growth rate were subsequently validated via numerical simulations~\cite{samtaney_formation_2009,ni_linear_2010,loureiro_plasmoid_2013}. 

In a non-rigorous, but rather convenient, fashion, the plasmoid-instability scalings can be easily retrieved from the usual tearing-mode expressions by replacing the scale-length characteristic of the background, tearing-unstable, equilibrium, $a$, with the thickness of the SP sheet, i.e., $a\rightarrow \delta_{SP}\sim L S^{-1/2}$~\cite{bhattacharjee_fast_2009,loureiro_plasmoid_2013}\footnote{It has recently been brought to our attention by K. Shibata that this particular way of deriving the plasmoid instability scalings was already known quite some time ago --- see problem 3-6 of Ref.~\cite{tajima_plasma_2002}.}. 
We note that it is not {\it a priori} obvious that this procedure should yield correct results because an SP sheet differs from the equilibria considered in the standard tearing analysis in that it contains sheared differential inflows and outflows. 
However, the fact that the growth rate of the plasmoid instability is super-Alfv\'enic justifies the neglect of such terms~\cite{loureiro_instability_2007,loureiro_plasmoid_2013}.
This useful observation allows the straightforward derivation of the plasmoid instability scalings in other plasma regimes of interest. For example, the large magnetic Prandtl number case is treated in Ref.~\cite{loureiro_plasmoid_2013}; and, in \secref{sec:kin_plasmoids} of this paper, we obtain scalings for semi-kinetic regimes in a similar fashion.

The original analysis of the plasmoid instability~\cite{loureiro_instability_2007} was essentially one-dimensional, in the sense that it did not address the effects of the dependence of the upstream magnetic field on the outflow coordinate, i.e., the theory rigorously applied only to a restricted vicinity of the centre of the sheet. It also did not allow for any variation in the out-of-plane direction.
With regards to the former limitation, Ref.~\cite{loureiro_instability_2007} generalised the analysis to a fully 2D SP-like equilibrium~\cite{loureiro_plasmoid_2013}, with a perhaps surprising result. While in the central part of the sheet the $S$-dependence of the plasmoid instability is not affected, they find that the wave-number and growth rate of fastest growing mode actually {\it increase} along the sheet and, remarkably, that there is a location (the Alfv\'en Mach point) where the plasmoid instability is replaced by the Kelvin-Helmholtz (KH) instability. To the best of our knowledge, this prediction has not yet been confirmed in direct numerical simulations (we speculate that this could be because triggering the KH instability may require even higher values of the Lundquist number than the $S_c\sim10^4$ required for the plasmoid instability).

With regard to 3D effects, these have been considered by Baalrud~\etal~\cite{baalrud_reduced_2012}, who conclude that the most unstable modes are oblique, i.e., align themselves at an angle between the reconnecting field and the out-of-plane (guide-) field (though the $S$ dependence of the scalings is not affected by this). 

\subsection{Nonlinear stage}
\label{sec:nonlinear_stage}

The linear theory of the plasmoid instability is valid for as long as $w \ll \delta_{in}\sim \delta_{SP}S^{-1/8}$, where $w$ is the width of the plasmoid chain~\cite{loureiro_instability_2007,loureiro_plasmoid_2013}.
However, in order to affect the reconnection process significantly, plasmoids have to become wider than the original SP sheet. 
In other words, understanding the effect of plasmoids on reconnection requires understanding their nonlinear evolution. 
This can be expected to be rather complex, even chaotic~\cite{lapenta_self-feeding_2008}. 
In principle, nonlinear plasmoid dynamics should be determined by the balance between the following processes: (i) nonlinear growth via reconnected flux accumulating in the plasmoids, (ii) advection along and ejection out of the sheet by the large-scale, background, sheared, Alfv\'enic reconnection outflows, (iii) coalescence (mergers of plasmoids with each other) and (iv) plasmoid saturation. 
Secondary current sheets are expected to form between neighbouring plasmoids; these current sheets will themselves be susceptible to the plasmoid instability if the local Lundquist number (defined for each sheet in terms of its length) exceeds $S_c$. 
This gives rise to a hierarchical, fractal-like structure~\cite{shibata_plasmoid-induced-reconnection_2001} which ends when the length of the inter-plasmoid current sheet is such that the local Lundquist number is $\sim S_c$.
Thus, at any given moment in time, one should expect the reconnection layer to be a stochastic plasmoid chain, with a distribution of plasmoids of different sizes and fluxes, and a total number of plasmoids given by $N\sim L/L_c$ (cf.~\cite{cassak_scaling_2009}).  

A statistical theory including all these different ingredients was proposed by Uzdensky~\etal~\cite{uzdensky_fast_2010}. It made three main predictions: (i) the effective (time-averaged) reconnection rate is determined by the SP model applied to the current sheets at the bottom of the plasmoid hierarchy, i.e., $\tilde E_{\rm eff}\equiv cE_{\rm eff}/(V_A B_0)\sim S_c^{1/2}$ --- note that this is {\it fast} since it does not depend on the global Lundquist number; (ii) the plasmoid size and flux distribution functions are, respectively, $f(w)\propto w^{-2}$ and $f(\psi)\propto\psi^{-2}$; (iii) abnormally large (monster) plasmoids should form occasionally, with widths roughly $w\sim \tilde E_{\rm eff}^{1/2}L$.

These estimates are in agreement with the results of high-Lundquist number simulations~\cite{bhattacharjee_fast_2009,huang_scaling_2010,loureiro_magnetic_2012}, although the distribution functions have been found to be more complex than originally thought: while the numerical data is in good agreement with the predicted $-2$ slope of the distribution function for plasmoids exceeding a certain flux/width threshold, below that threshold the slope is shallower than $-2$, and is instead consistent with a $-1$ power law index~\cite{huang_distribution_2012}. Explanations for this transition in the spectra have been put forth~\cite{loureiro_magnetic_2012,huang_distribution_2012}. 
A detailed discussion of this issue is beyond the scope of this paper; it is however worth adding that a firm grasp of the expected plasmoid distribution function is directly relevant to the problem of particle acceleration in reconnection~\cite{drake_electron_2006,chen_observation_2008,drake_power-law_2013}, although this requires going beyond the MHD description, as does the detailed interpretation of observations~\cite{nishizuka_multiple_2010, fermo_comparison_2011}. Other physical contexts in which the details of the plasmoid distribution function may matter are discussed in~\secref{sec:HED}.

Another point worth making concerns the prediction of the occasional formation of monster plasmoids.  
In essence, their existence requires a reconnecting plasmoid chain that has a flow stagnation point close to its geometric center. We note that this prescription is not exclusive to MHD, i.e., fully kinetic (PIC) simulations of reconnection share this property \cite{drake_electron_2006,daughton_role_2011,liu_dispersive_2014}. Thus, we see no {\it a priori} reason why monsters should not also exist in kinetic environments (with an even larger size, since the reconnection rate in a kinetic plasmoid chain is roughly a factor of $10$ larger than its MHD counterpart, on the basis of the available numerical evidence).

Finally, we note that the plasmoid instability provides a very natural way to trigger a transition to kinetic physics in reconnecting systems where, on the basis of the SP model, one would previously have predicted MHD to be a valid description. Indeed, even if ion kinetic scales are smaller than $\delta_{SP}$, it may still be the case that they exceed the width of the critical sheet, $\delta_c\sim L_c S_c^{-1/2}$, at which point a transition to kinetic reconnection will be triggered~\cite{daughton_transition_2009,daughton_influence_2009,shepherd_comparison_2010,uzdensky_fast_2010,huang_onset_2011}.

\section{The plasmoid instability in context}
\label{sec:plasmoid_context}
Over the last few years, the plasmoid instability has been discussed in a wide variety of contexts, ranging from magnetically confined laboratory plasmas to very diverse space and astrophysical phenomena.
One of its most direct applications has been solar flares, where there are many observational reports of plasmoid-like features (e.g.,~\cite{nishizuka_multiple_2010, karlicky_reconnection_2010, barta_spontaneous_II_2011,takasao_simultaneous_2012}; see~\cite{loureiro_plasmoid_2013} for further references in this and other contexts), supported by direct numerical simulations (e.g.,~\cite{barta_spontaneous_2011,karpen_mechanisms_2012,murphy_asymmetric_2015}).
The aim of this section is to underscore the fundamental role played by plasmoids in a few other applications where this is not yet, perhaps, as widely appreciated.

\subsection{Plasmoids in tokamaks}
\label{sec:plasmoids_tokamaks}

Magnetically confined fusion experiments --- tokamaks in particular, but not only --- are prone to a variety of instabilities where reconnection plays a fundamental role: the sawtooth and (classic, neoclassic and micro) tearing  modes, resonant magnetic perturbations and (probably) edge localized modes (ELMs) are a few important examples. 
Recent research has underscored the importance of plasmoids in two of these phenomena: the sawtooth instability and resonant magnetic perturbations. 
We briefly describe these results in the following paragraphs.

\subsubsection{Sawtooth Instability.}
\label{sec:sawtooth}
The periodic relaxation of the temperature in the core of tokamaks --- which results in a temporal evolution of the core temperature displaying a characteristic sawtooth-like pattern --- is one of the earliest observed instabilities in such devices~\cite{hastie_sawtooth_1997, yamada_magnetic_2010}. A very elegant attempt to describe it was proposed by Kadomtsev~\cite{kadomtsev_disruptive_1975}. His explanation involves the following key steps: (i) an externally imposed electric field drives the toroidal current in the plasma, which in turn Ohmically heats the plasma; (ii) as the electron temperature ($T_e$) thus increases, the plasma resistivity ($\eta$) decreases ($\eta\propto T_e^{-3/2}$); (iii) to maintain the externally imposed electric field, Ohm's law $E=\eta j$ requires that the current in the plasma increases to compensate for the decrease in resistivity; (iv) raising the current leads to a lowering of the safety profile ($q$) in the core; (v) if $q$ in the core becomes less than one, an internal kink mode is triggered; this then drives reconnection at the $q=1$ surface; (vi) reconnection at this surface proceeds according to the Sweet-Parker model, the hot core plasma mixes with the colder plasma found at radial locations $r>r_1$ (where $r_1$ is the radial location of the $q=1$ surface), resulting in a lower value of $T_e$ in the (new) core; (vii) the cycle repeats for as long the external electric field is on, resulting in the sawtooth pattern.

Sadly, the understanding gathered over the years, theoretically, computationally and experimentally, has established beyond doubt that Kadomtsev's model provides an insufficient explanation of the sawtooth instability (and may even be incompatible with some observations)~\cite{hastie_sawtooth_1997, yamada_magnetic_2010}. 
Perhaps not surprisingly, the first problem that was identified was the model's inability to reproduce the timescale of the crash phase, identified with the reconnection stage in the model. This is simply an extension to the sawtooth instability of the problem faced by the SP model in almost all other contexts to which one choses to apply it: SP reconnection is too slow (taking JET as a specific example, sawtooth crash times are $\sim 100 \mu$s, whereas Kadomtsev's model yields $\sim 10$ms).

In view of the foregoing discussion on the plasmoid instability, it is a straightforward conjecture that the SP-like current layers envisaged in Kadomsev's sawtooth model (which form in the nonlinear evolution of the kink mode) may in fact not be stable --- one may legitemately expect them to be susceptible to the plasmoid instability, break into many plasmoids and undergo fast (independent of $S$) reconnection.
Very recent computational work by Yu \etal~\cite{yu_formation_2014} has shown exactly this. 
Of course, it has to be added that MHD is an insufficient description of the sawtooth cycle in modern tokamaks and that, while the sawtooth crash may involve plasmoid formation~\cite{gunter_fast_2015}, its full understanding remains a mystery.

\subsubsection{Error fields and resonant magnetic perturbations: the Taylor problem.}
\label{sec:Taylor}

A well-known paradigm to investigate error fields (small amplitude deviations from the idealized magnetic equilibria) and resonant magnetic perturbations in tokamaks was proposed by J. B. Taylor~\cite{hahm_forced_1985}; the basic idea is to investigate how small-amplitude perturbations imposed far away from a tearing-stable rational layer drive reconnection at that layer (hence the term forced, or driven, to characterize the reconnection events that arise in this way)  --- we refer the interested reader to Ref.~\cite{comisso_phase_2015} for a clear description of the problem and overview of relevant work in this topic.

Until very recently, the accepted theoretical understanding of what came to be known as the Taylor problem comprised two different regimes: (i) Hahm-Kuslrud~\cite{hahm_forced_1985} and (ii) Wang-Bhattacharjee~\cite{wang_forced_1992}. In the former, reconnection is very slow --- the nonlinear stage is well-described by Rutherford's theory~\cite{rutherford_nonlinear_1973}; in the latter, a Sweet-Parker current sheet forms in the nonlinear regime~\cite{waelbroeck_current_1989, loureiro_x-point_2005} and, consequently, reconnection is faster (but, of course, not fast). Which scenario is followed by a given perturbation is essentially a function of the perturbation's amplitude: smaller perturbations conform to the theory of Hahm and Kulsrud, larger perturbations to that of Wang and Bhattacharjee.

In view of what has been learned about the unforced case, it could be expected that a third regime may exist --- one in which the nonlinear current sheet, as it is formed, becomes unstable to plasmoids, thus replacing the Wang-Bhattacharjee scenario with one governed by plasmoid-induced fast reconnection. 

This possibility has indeed been posited in recent analytical work by Dewar \etal~\cite{dewar_plasmoid_2013} and subsequently demonstrated in numerical work by Comisso and co-workers~\cite{comisso_extended_2015,comisso_phase_2015}.
The calculations by these authors suggest that the perturbation amplitudes required to reach this new scenario are small enough to be experimentally relevant.

\subsection{Plasmoids in high-energy-density astrophysical phenomena}
\label{sec:HED}

Several important astrophysical phenomena, most notably the magnetospheres of magnetars and central engines of gamma-ray bursts (GRBs) and supernovae, take place in environments with such a high energy density and, in particular, with such a strong magnetic field, that dissipation of the magnetic energy via reconnection inevitably leads to intense prompt pair creation and thus renders  the plasma highly collisional~\cite{uzdensky_stellar_2006,uzdensky_magnetic_2011,mckinney_reconnection_2012} (see Ref.~\cite{uzdensky_plasma_2014} for a review).  This, in turn, means that any magnetic reconnection processes that might take place in such environments should proceed in the resistive MHD regime.  
However, the estimated Lundquist numbers in these systems are so huge that Sweet-Parker reconnection would be hopelessly slow. 
At the same time, fast magnetic reconnection has been conjectured to play an important role in these systems; in particular, it is a leading mechanism for explaining the giant gamma-ray flares in magnetar systems like Soft Gamma Repeaters (e.g., ~\cite{lyutikov_explosive_2003,lyutikov_magnetar_2006,uzdensky_magnetic_2011}), and has also been proposed as the main mechanism powering the prompt gamma-ray emission in GRBs~\cite{drenkhahn_efficient_2002,giannios_spectra_2005,lyutikov_electromagnetic_2006,mckinney_reconnection_2012}. 
It is only thanks to the plasmoid-dominated reconnection regime, with dimensionless reconnection rates of 0.01 or faster, that fast reconnection is possible in such systems. 

Furthermore, the inherently non-steady, bursty character of energy dissipation in plasmoid-mediated reconnection, especially when combined with the kinetic beaming effect~\cite{cerutti_beaming_2012}, provides a natural explanation for the ultra-rapid, multi-scale time variability of the high-energy emission observed in many flaring astrophysical systems, such as TeV flares in blazar jets~\cite{nalewajko_energetic_2012,giannios_reconnection-driven_2013}, gamma-ray flares in the Crab pulsar wind nebula~\cite{cerutti_beaming_2012,cerutti_simulations_2013,cerutti_three-dimensional_2014}, and GRBs~\cite{uzdensky_magnetar-driven_2007,uzdensky_plasma_2014}.  

\section{Outstanding questions}
\label{sec:outstanding}

We now wish to discuss briefly a few selected topics that appear to us to be a natural and necessary continuation of research in plasmoid-dominated reconnection.

\subsection{Parametric dependence of the critical Lundquist number}
\label{sec:Scrit}
A central role in the theoretical description of reconnection in the plasmoid regime is played by the critical Lundquist number, $S_c$; this determines (i) whether a pre-formed current sheet is unstable to the plasmoid instability, (ii) the dimensions of the current sheets found at the bottom of the plasmoid hierarchy ($\delta_c/L_c\sim S_c^{-1/2}$) and (iii) the reconnection rate ($c E_{\rm eff}\sim S_c^{-1/2}V_AB_0$).

The theory of the plasmoid instability~\cite{loureiro_instability_2007, loureiro_plasmoid_2013} cannot be used to predict $S_c$ rigorously  because it is an asymptotic theory, i.e., it assumes that $S \gg S_c$. A non-rigorous extrapolation can however be made~\cite{loureiro_plasmoid_2013} based on the requirement of reasonable scale separation between the boundary layer of linear theory\footnote{As in the usual tearing mode calculation~\cite{furth_finite-resistivity_1963}, the linear theory of the plasmoid instability~\cite{loureiro_instability_2007, loureiro_plasmoid_2013} divides the domain into the immediate vicinity of the rational layer (the boundary layer), where resistive effects on the perturbation are important, and regions away from the rational layer, where resistivity can be neglected.} and the thickness of the current sheet, i.e., $\delta_{in}/\delta_{CS}\sim S^{-1/8}\ll 1$; in rough terms, a minimum requirement for the validity of asymptotic theory is that $\delta_{in}/\delta_{CS}\sim 1/3$, which suggests $S_c\sim 10^4$ (similar arguments can be drawn based on the growth rate and wave number of the fastest growing mode; these, however, yield less stringent requirements on $S$ than $\delta_{in}$ since the latter has the weakest dependence on the Lundquist number). 
The values of $S$ typically reported in the numerical literature tend to range from a few thousand to $\sim 10^4~$\cite{biskamp_magnetic_1986,loureiro_x-point_2005,samtaney_formation_2009,cassak_scaling_2009,loureiro_magnetic_2012}.

At any rate, $S_c$ is set by tearing-mode dynamics in a current channel with sheared plasma flows; the precise marginal stability boundary of this system will naturally be a function of plasma parameters such as  $\beta$~\cite{ni_effects_2012,baty_effect_2014}, viscosity~\cite{loureiro_plasmoid_2013}, relativistic effects~\cite{takamoto_evolution_2013}, etc. 
The fact that reconnection happens in a wide variety of environments, where some of these parameters can take extremely different values, suggests that a firmer grasp of this dependence may be quite important and actually lead to appreciable differences in reconnection rates. 

\subsection{Onset}
\label{sec:onset}

The fact that large-aspect ratio, SP-like current sheets are super-critical systems prompts one to reassess how they may form in the first place ~\cite{uzdensky_magnetic_2014,pucci_reconnection_2014, tenerani_self-similar_2015}. In other words, starting from a current-sheet-free plasma, one imagines that there is some continuous dynamical process (e.g., turbulence, but the exact mechanism is not relevant at this stage) that leads to current sheet formation. In the course of this, the aspect ratio of the forming current sheet increases as a function of time and thus one expects that the current sheet progressively approaches the  marginal stability boundary of the plasmoid instability. 
As this happens, very little reconnection occurs, i.e., this corresponds to a slow stage of energy accumulation. However, as the current-sheet aspect ratio continues to increase, the marginal-stability threshold will eventually be crossed, leading to the onset of the plasmoid instability, and the consequent transition to fast reconnection soon thereafter. This sequence of steps would naturally preclude the formation of a fully-developed SP sheet of aspect ratio $\delta_{SP}/L\sim S^{-1/2}$. 
Understanding the formation and evolution of plasmoids in a {\it forming} current sheet is an important open question in reconnection research; in particular, the moment of time when a forming current sheet is disrupted by plasmoids may be closely related to the {\it reconnection trigger}, or {\it onset}~\cite{uzdensky_magnetic_2014}.

\subsection{3D}
\label{sec:3D}
Much of what has been learned about magnetic reconnection so far stems from relatively simple two-dimensional configurations, and it would appear that our knowledge of this subject has now reached a certain level of maturity --- certainly as far as MHD is concerned, but perhaps even beyond that.
In tandem with the computing capabilities that are available with today's best super-computers, it thus seems that the time is ripe for tackling what we strongly believe to be the next research frontier in reconnection: fully 3D geometries.

Several numerical works have appeared recently in the literature suggesting that 3D MHD reconnection may be quite different than 2D, with self-excited turbulence (of both the plasmoid and non-plasmoid kind) leading to rather complex configurations~\cite{lapenta_spontaneous_2011,beresnyak_rate_2013,lazarian_turbulent_2015, oishi_self-generated_2015} (state-of-the-art 3D kinetic simulations reveal qualitatively similar, but of course physically even richer, behaviour~\cite{daughton_role_2011}.)
Matching experimental evidence for such complexity has also been reported~\cite{intrator_flux_2013,gekelman_chaos_2014}.
Simultaneously, a series of analytical papers by Boozer~\cite{boozer_reconnection_2002,boozer_magnetic_2012,boozer_separation_2012, boozer_model_2013, boozer_formation_2014} has presented very compelling arguments for an altogether different paradigm for 3D reconnection, where the (probably unavoidable in space and astrophysical plasmas) exponentiation of the distance between two neighbouring field lines may lead to fast reconnection at very low levels of the current intensity (cf. Ref.~\cite{huang_rapid_2014}).

\begin{figure}
\begin{center}
\includegraphics[width=0.5\textwidth]{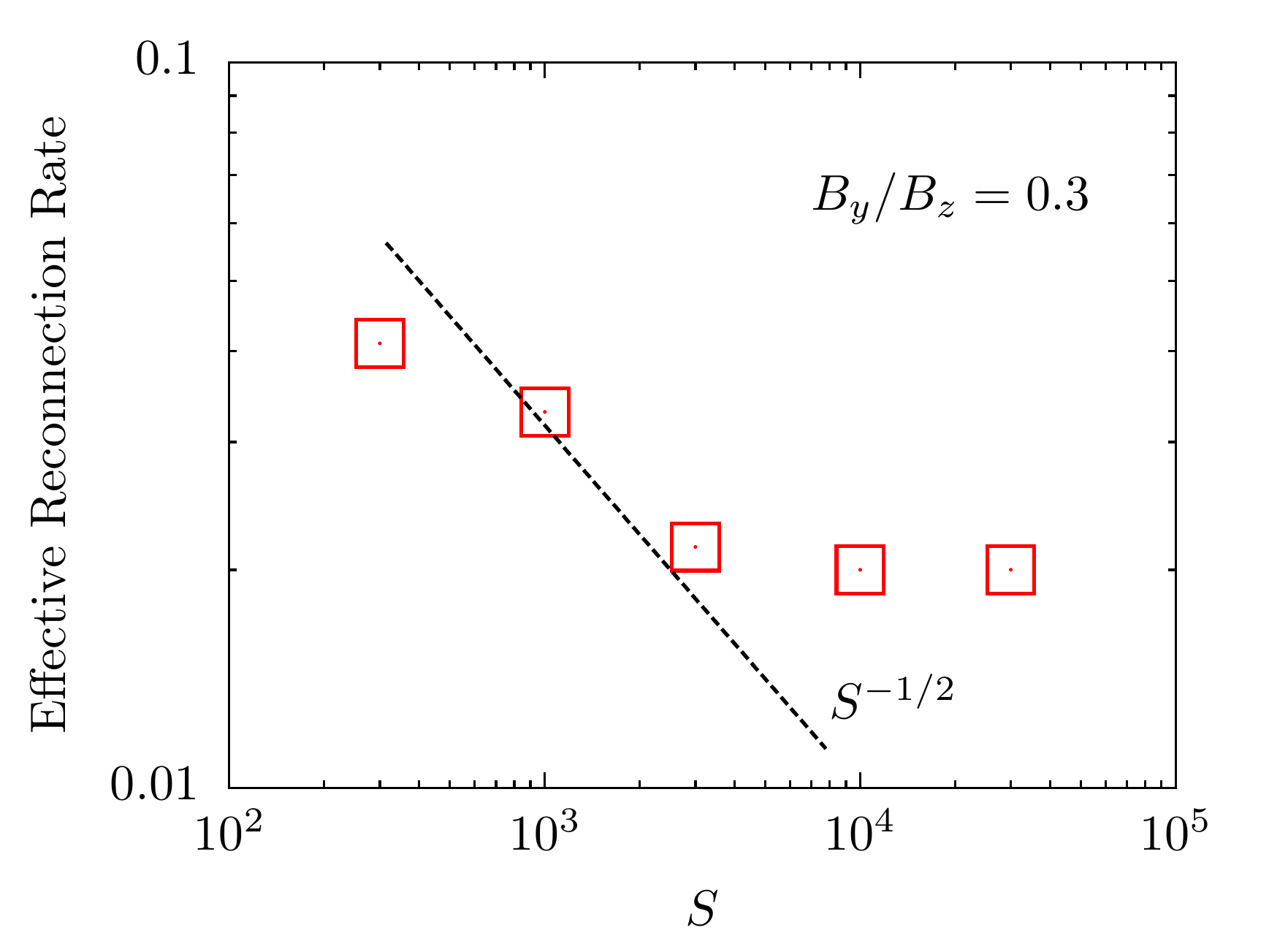}
\caption{The effective reconnection rate as a function of the Lundquist number in 3D MHD simulations~\cite{schoeffler_3D_2015}. In this case, $B_y/B_z=0.3$, where $B_y$ is the amplitude of the reconnecting magnetic field, and $B_z$ is the guide magnetic field (i.e., the component perpendicular to the reconnection plane).}
\label{fig:3D_rec_rate}
\end{center}
\end{figure}

While a detailed understanding of magnetic reconnection in intrinsically 3D geometries remains an open  challenge, a straightforward step in that direction is the extension to 3D of previous 2D studies of the plasmoid instability~\cite{loureiro_magnetic_2012}. 
As a preliminary result, we
plot in~\fig{fig:3D_rec_rate} the effective reconnection rate (measured as the ratio between the time-averaged plasma inflow and outflow velocities) as a function of the Lundquist number. 
The simulations are performed in a 3D ``semi-global'' slab where the $x$- and $y$-directions define the reconnection plane (the inflow and outflow directions). 
The dimensions of the simulation domain are $L_y=L_z=0.5 L$, where $L$ is the (arbitrary) system size, and $L_x$ depends on the Lundquist number but is always $\gtrsim 10 \delta_{SP}$. 
We specify periodic boundary conditions in the $z$-direction, free-outflow boundary conditions in $y$ and impose the density, pressure and magnetic field at the $x$ boundaries (see~\cite{loureiro_magnetic_2012} for details).
The scan in $S$ is performed at fixed $B_y/B_z=0.3$, where $B_y$ is the magnitude of the upstream (reconnecting) magnetic field, and $B_z$ is the guide magnetic field. The fluid viscosity is always equal to the resistivity.

The reconnection rates shown in~\fig{fig:3D_rec_rate} exhibit the same trend as found in 2D simulations~\cite{lapenta_self-feeding_2008, loureiro_turbulent_2009,bhattacharjee_fast_2009,cassak_scaling_2009, huang_scaling_2010,loureiro_magnetic_2012}: at moderately small values of $S$, the Sweet-Parker $S^{-1/2}$ scaling holds, followed by a transition to an $S$-independent reconnection rate of $\sim 0.02$ (same as found in our previous 2D simulations~\cite{loureiro_magnetic_2012}) as $S$ increases beyond a critical value $\sim 3\times 10^3$ (whereas in the 2D case we found $S_c\sim 10^4$, but see the discussion below). 
This transition is accompanied by plasmoid (flux rope) formation (not shown); the number of plasmoids observed increases with $S$. 

These observations strongly suggest that the 2D statements about the plasmoid instability rendering the reconnection rate independent of $S$ are robust. 
However, there are several important questions that have not yet been addressed: does the transition to the plasmoid-stage, and the reconnection rate found therein, depend on the strength of the guide field? What is the structure and the distribution function of 3D plasmoids (flux-ropes)? How is the energy balance changed from the 2D case? Ref.~\cite{loureiro_magnetic_2012} shows that, in statistical steady-state, roughly $40\%$ of the incoming magnetic energy is dissipated via Ohmic and viscous heating; interestingly, Beresnyak~\cite{beresnyak_rate_2013} quotes the same number in his 3D simulations  --- how does this fraction depend on the physical parameters of the plasma? In particular, is reconnection always an efficient energy dissipation mechanism? A study addressing some of these issues, as well as particle acceleration in 3D, plasmoid-dominated, MHD reconnection environments is currently underway~\cite{schoeffler_3D_2015}.

\subsection{Plasmoids beyond the MHD description}
\label{sec:kin_plasmoids}
So far, our discussion has been limited to reconnection and plasmoids in MHD plasmas. 
This obviously excludes many reconnecting regimes where there is abundant evidence of plasmoid formation~\cite{ji_phase_2011} but where collisions are not sufficiently frequent (compared to the timescales of interest) to justify the use of an MHD description.
We next discuss two such scenarios, in order of decreasing plasma collisionality. 

\subsubsection{The semi-collisional plasmoid instability.}
\label{sec:semi-collisional}
The semi-collisional regime of the plasmoid instability is defined by 
\be
\label{eq:semi_coll_plasmoids}
\delta_{CS}\gg \lambda_i\gg \delta_{in}\gg \lambda_e,
\ee
where $\lambda_j$, with $j=i,e$, denotes the ion or electron kinetic scales most relevant to the particular regime of interest, i.e., $\rho_i,\rho_s,$ or $d_i$ in the case of the ions, $d_e$ or $\rho_e$ in the case of the electrons, with $d_j=c/\omega_{pj}$ the ion/electron skin-depth, $\rho_{j}$ the ion/electron Larmor radius and $\rho_s$ the ion sound Larmor radius, and $\delta_{in}\sim \delta_{SP}S^{-1/8}$ the width of the boundary layer that arises in the linear MHD plasmoid instability analysis (see \secref{sec:linear_stage}).

This regime is amenable to analysis via direct application of the standard tools of the tearing instability theory. 
The case when $\lambda_i=d_i$ [Hall-MHD plasmoid regime, relevant when the magnetic-field component perpendicular to the reconnection plane (the guide-field) is weak (i.e., $\beta\gg 1$)] has been rigorously analysed by Baalrud \etal~\cite{baalrud_hall_2011}. They found $\gamma_{\rm max} \tau_A\sim (d_i/L)^{16/3}S^{7/13}$ and $k_{\rm max}  L=(d_i/L)^{1/13}S^{11/26}$ for the most unstable mode. 

In the opposite limit of strong guide field ($\beta\ll 1$) one instead needs to consider $\lambda=\rho_s$.
The plasmoid instability scalings in this case can be easily derived (non-rigorously, as described in~\secref{sec:linear_stage}) from well-known results for the tearing instability in the semi-collisional regime (see~\cite{zocco_reduced_2011}, Appendix 5c., and references therein; we add for completeness that the scalings above for $\lambda_i=d_i$ can be retrieved in a similar way from the results of Ref.~\cite{fitzpatrick_collisionless_2004}). 
We obtain:
\bea
\label{eq:semi_collisionless_plasmoid_gmax}
\gmax^{\rm SC} L/V_A\sim  (\rho_s/L)^{2/3}S^{2/3},\\
\label{eq:semi_collisionless_plasmoid_kmax}
\kmax^{\rm SC} L \sim (\rho_s/L)^{1/9} S^{4/9}, \\ 
\label{eq:semi_collisionless_plasmoid_deltain}
\delta_{in}^{\rm SC}/L\sim \(\rho_s/L\)^{-7/9}S^{-10/9},
\eea
where the upper subscript SC denotes `semi-collisional' to avoid confusion with the corresponding scalings in the pure MHD regime. These scalings are qualitatively similar to the MHD ones, in the sense that the instability grows faster at higher $S$, and the wave-number increases, which again leads to the conclusion that Sweet-Parker current sheets in plasmas satisfying~\eq{eq:semi_coll_plasmoids} cannot form in the first place, because they are violently unstable.

In addition to the constraints expressed in \eq{eq:semi_coll_plasmoids}, the instability requires that all of the following conditions be satisfied: $\gmax^{\rm SC} L/V_A\gg1$, $\kmax^{\rm SC} L\gg 1$ and $\delta_{in}^{\rm SC}/\rho_s\ll 1$. The latter is the most stringent, requiring $S\gg (L/\rho_s)^{8/5}$, the same threshold that is yielded by the second inequality in \eq{eq:semi_coll_plasmoids}; thus, the critical value of the Lundquist number is 
\be
\label{eq:SC_S_crit}
S_c^{\rm SC}\sim (L/\rho_s)^{8/5},
\ee
but $S\ll (L/\rho_s)^2$, as required by the first inequality in \eq{eq:semi_coll_plasmoids} (or else $\rho_s\gg\delta_{CS}$ and it no longer makes sense to consider a Sweet-Parker current sheet as the background equilibrium)\footnote{Baalrud \etal~\cite{baalrud_hall_2011} retrieve the same criteria in the case of the Hall-MHD regime for $L/d_i$ instead of $L/\rho_s$, but further require that this be an additional constraint besides the MHD one, i.e., that in addition to $S\gg (L/d_i)^{8/5}$, the Lunqduist number must also be such that $S>S_c\sim 10^4$. As we explained earlier (and also in~\cite{loureiro_plasmoid_2013}), $S_c\sim 10^4$ stems from the requirement that the boundary layer for the resistive-MHD version of the plasmoid instability fit inside the current sheet. In the semi-collisionless regime that we are addressing in this section, the MHD boundary layer is replaced with one whose width is given by~\eq{eq:semi_collisionless_plasmoid_deltain}, so $S_c\sim10^4$ no longer follows from anywhere and must be replaced simply with \eq{eq:SC_S_crit}, or $S_c\sim (L/d_i)^{8/5}$ in the higher-$\beta$ case analysed by  Baalrud.}.

It is worth emphasizing that in the semi-collisional regime $S_c$ is {\it a function} of $L/\rho_s$, whereas in the MHD regime it is {\it a number} --- an evident reflection of the absence of special scales in MHD. This has interesting consequences that we now discuss. 

\begin{figure}
\begin{center}
\includegraphics[width=0.8\textwidth]{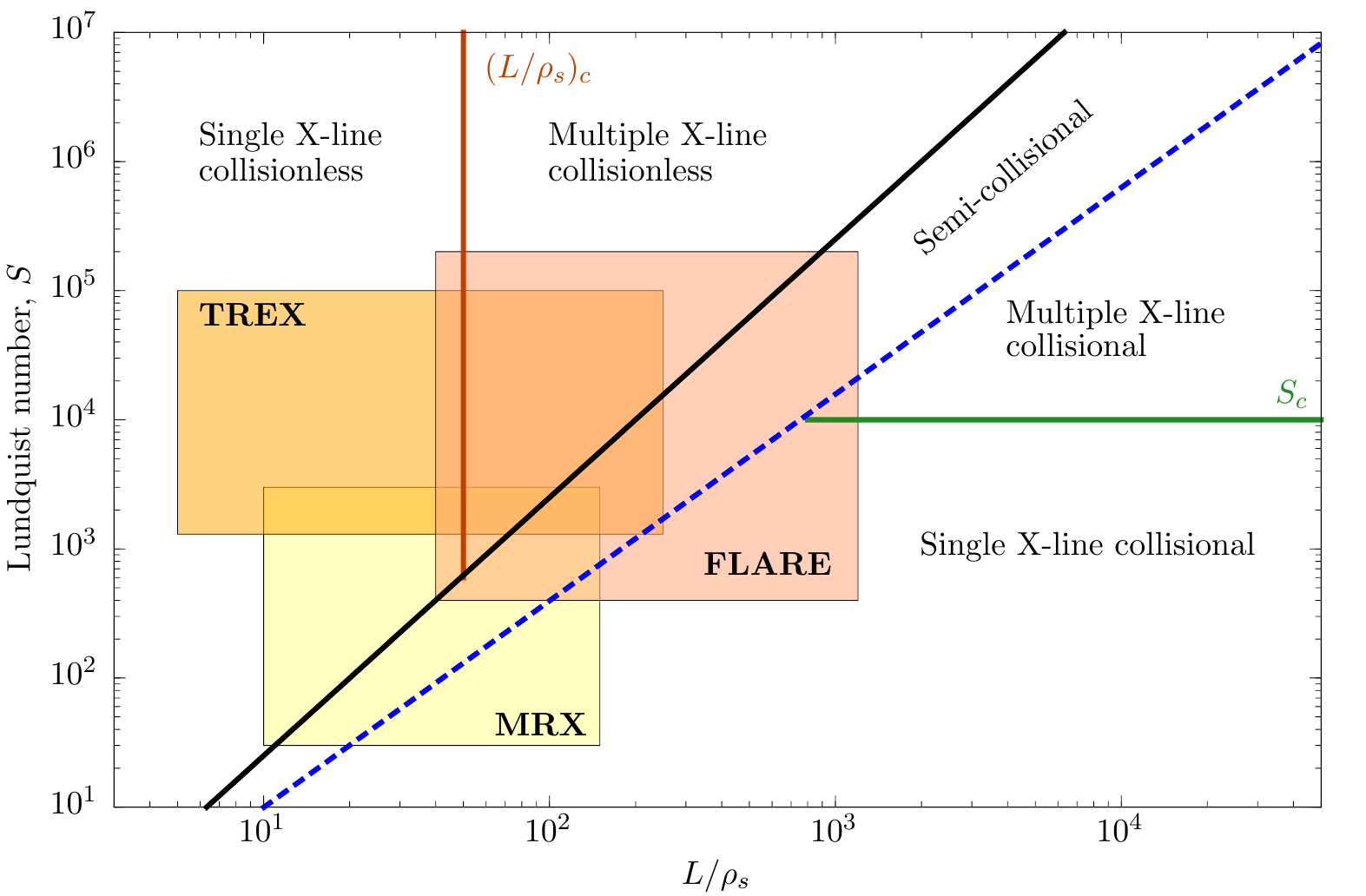}
\caption{Reconnection phase diagram (cf.~\cite{ji_phase_2011}). The semi-collisional plasmoid theory suggests the existence of a new reconnection regime that may be pertinent to past and future reconnection experiments. (See~\secref{sec:semi-collisional} for an explanation of the different lines and regimes indicated in the figure).}
\label{fig:diagram}
\end{center}
\end{figure}

In \fig{fig:diagram} we revisit the reconnection phase diagram of Ji and Daughton~\cite{ji_phase_2011} (see also~\cite{huang_plasmoid_2013}). 
Different reconnection regimes are indicated as a function of Lundquist number (vertical axis) and of the ratio between the system size, $L$, and the ion sound Larmor radius, $\rho_s$ (horizontal axis). The (black, diagonal) solid line is yielded by the Sweet-Parker model (and forgetting for the moment the plasmoid instability): comparing $\delta_{SP}$ with kinetic effects (whose proxy here is $\rho_s$) indicates whether we are in a collisional SP regime, or in a collisionless regime (respectively, below or above the solid black line).
The vertical (solid) red line labelled $(L/\rho_s)_c$ is an empirical line that follows from the numerical observation that simulations of collisionless reconnection with $L/\rho_s\gtrsim 50$ tend to exhibit multiple plasmoids, whereas those with $L/\rho_s\lesssim 50$ tend to show a single X-point (there is no analytical theory to back this threshold).
The horizontal (solid) green line labelled $S_c$ indicates the resistive MHD plasmoid instability threshold --- the SP current sheet is plasmoid unstable in the region above the green line, and stable below it (another threshold that exists but that we omit for simplicity results from the case when the MHD plasmoid instability, in its nonlinear stage, triggers a transition to the kinetic scales~\cite{uzdensky_fast_2010}). The regions of operational space roughly covered by  a selection of past (MRX~\cite{yamada_study_1997,ji_experimental_1998}), present (TREX~\cite{trex}) and future (FLARE~\cite{flare}) reconnection experiments are also indicated.
Finally, we draw in dashed blue the new line suggested by the semi-collisional theory discussed above: plasmas (asymptotically) above this line, but (asymptotically) below the (black) solid diagonal line, should exhibit the semi-collisional version of the plasmoid instability. 

It is worth contrasting this diagram with its previous version~\cite{ji_phase_2011} (remove the blue dashed line and extend the green horizontal line all the way to meet the black solid line) and realising that, in light of these new results,  the plasmoid instability, in the semi-collisional version described here (or in the version of Ref.~\cite{baalrud_hall_2011} if $\beta$ is not small), should actually be accessible to existing facilities such as MRX, and even more accessible to new and upcoming experiments like TREX and FLARE. 

In addition, we remark that another important reconnection environment that falls in the semi-collisional regime is the solar corona (not shown in the diagram because it is truncated at relatively low values of $S$).

A nonlinear theory of the semi-collisional plasmoid instability is an open (and in view of this discussion, important) problem.

\subsubsection{The semi-collisionless plasmoid instability.}
At somewhat smaller values of collisionality, the ordering of the relevant length scales becomes
\be
\label{eq:semi_coll-less_plasmoids}
\delta_{CS}\gg \lambda_i \gg \lambda_e\gg \delta_{in}.
\ee
We shall refer to this as the semi-collisionless regime; essentially, it differs from the semi-collisional regime in that the breaking of the frozen-flux condition is now enabled by electron kinetic effects, not collisionality (even if, of course, collisionality must remain finite or there would not be a Sweet-Parker sheet in the first place).

Considering the case when $\lambda_i=\rho_s$ and $\lambda_e=d_e$ (and negligible electron finite-Larmor-radius effects), we derive from the corresponding scalings for the tearing instability (see Appendix 3.e of Ref.~\cite{zocco_reduced_2011} and references therein) the following expressions:
\bea
\gmax^{\rm semi-c'less}\sim (V_A/L)\,(d_e/L)\, (\rho_s/L)\, S^{3/2},\\
\kmax^{\rm semi-c'less}\sim L^{-1} (d_e/L)^{2/3}(\rho_s/L)^{1/3} S,\\
\delta_{in}^{\rm semi-c'less}\sim L (d_e/L)^{2/3} (\rho_s/L)^{1/3}.
\eea
Once again, the expressions for the growth rate and wave-number diverge (even faster than in the MHD or semi-collisional regimes) as $S$ increases.

The semi-collisionless regime formally holds if $\delta_{in}^{\rm SC} \ll d_e$ and $\delta_{in}^{\rm semi-c'less} \ll \rho_s \ll \delta_{SP}$.
 The first of these requirements yields 
\be
\label{eq:S_requirements_semi_c'less}
S\gg (L/d_e)^{9/10} (L/\rho_s)^{7/10},
\ee 
whereas requiring that the $\delta_{in}^{\rm semi-c'less} \ll \rho_s$ demands
$d_e\ll \rho_s$, a condition already assumed in this derivation and which directly translates into $\beta_e\gg m_e/m_i$; 
finally, the requirement $\rho_s \ll \delta_{SP}$ was already dealt with in the semi-collisional case and translates into $S\ll (L/\rho_s)^2$. To appreciate the mutual consistency of these and the semi-collisional scalings above, note that taking $\rho_s = d_e$ converts~\eq{eq:S_requirements_semi_c'less} into $S\gg (L/\rho_s)^{8/5}$.  Thus, the critical value of $S$ required to access the semi-collisionless regime is obtained from \eq{eq:S_requirements_semi_c'less}, with maximum value of $(L/\rho_s)^2$.

\section{Final remarks and outlook}
\label{sec:conclusion}

In this paper, we have attempted to give a broad overview of recent developments and current state of the art in the study of magnetic reconnection in the simplest plasma description where reconnection is possible: resistive MHD. While there certainly are many examples of reconnecting environments where resistive MHD is an adequate description, it is also generally appreciated that there is a wide variety of weakly collisional systems where it is not. We would nonetheless argue that, even for such cases, a firmer grasp of MHD reconnection is critical: indeed, several MHD findings, and in particular the plasmoid instability and subsequent stochastic plasmoid dynamics, seem to carry over qualitatively to kinetic reconnection. 
In addition, we believe that the simple fact that MHD reconnection continues to surprise us justifies attempts to investigate it at a deeper level.

The developments in the field that we discussed here embolden us to venture the idea that the problem of magnetic reconnection in natural, high Lundquist number systems, may actually be a solvable one, in the sense of having concrete answers to the three overarching questions, viz., reconnection rate, trigger mechanism and energy partition. 
The suggestion that we may be nearing a satisfactory level of understanding of the 2D problem certainly seems credible, and indicates that the community's focus should perhaps turn to full 3D geometries. 
It is worth bearing in mind that, in contrast with another classic and fundamental problem of comparable complexity, namely, the turbulent dynamo, one may yet find that reconnection is {\it cursed}, rather than {\it blessed}, with being topologically possible in 2D. Indeed, the discovery that dynamos are mathematically impossible in 2D (Cowling's theorem) imposed realistic geometries on the dynamo community from very early on. The reconnection community has been able to get by so far in almost complete denial of 3D aspects, but this is changing as computers get ever more powerful. As this issue inevitably gains traction over the coming years, we will find whether existing, 2D, reconnection models translate to 3D in fairly obvious ways, or whether, on the contrary,  a completely different paradigm will emerge.

\section*{Acknowledgements}
The authors are grateful to A. A. Schekochihin for providing many suggestions that have considerably improved this paper. 
NFL was supported by the Funda\c{c}\~ao para a Ci\^encia e Tecnologia via
Grants UID/FIS/50010/2013 and IF/00530/2013.
\section*{References}
\providecommand{\newblock}{}

\end{document}